\documentclass[twoside,11pt]{article}

\usepackage{jmlr2e}
\usepackage{soul}
\usepackage{url}
\usepackage[utf8]{inputenc}
\usepackage[small]{caption}
\usepackage{graphicx}
\usepackage{amsmath}
\usepackage{booktabs}
\usepackage{algorithmic}
\usepackage{booktabs} 
\usepackage{amsmath,amssymb,amsfonts}
\usepackage{algorithmic}
\usepackage{graphicx}
\usepackage{textcomp}
\usepackage{xcolor}
\usepackage{float}
\usepackage{multirow}
\usepackage{array}
\usepackage{subfigure}
\usepackage{color}
\usepackage{subfigure}
\usepackage{fancyhdr}
\usepackage[ruled,linesnumbered,boxed]{algorithm2e}

\newtheorem{assumption}{Assumption}

\jmlrheading{Boyi Hou et al.}{2019}{}{}{}{}{}

\ShortHeadings{Gradual Machine Learning for Entity Resolution}{Hou and Chen}
\firstpageno{1}

\begin{document}

\title{Gradual Machine Learning for Entity Resolution}

\author{\name Boyi Hou \email ntoskrnl@mail.nwpu.edu.cn\\
       \addr School of Computer Science, Northwestern Polytechnical University
       \AND
       \name Qun Chen \email chenbenben@nwpu.edu.cn\\
       \addr School of Computer Science, Northwestern Polytechnical University
       \AND
       \name Yanyan Wang \email wangyanyan@mail.nwpu.edu.cn\\
       \addr School of Computer Science, Northwestern Polytechnical University
       \AND
       \name Youcef Nafa \email youcef.nafa@nwpu.edu.cn\\
       \addr School of Computer Science, Northwestern Polytechnical University
       \AND
       \name Zhanhuai Li \email lizhh@nwpu.edu.cn\\
       \addr School of Computer Science, Northwestern Polytechnical University}

\maketitle

\begin{abstract}
Usually considered as a classification problem, entity resolution (ER) can be very challenging on real data due to the prevalence of dirty values. The state-of-the-art solutions for ER were built on a variety of learning models (most notably deep neural networks), which require lots of accurately labeled training data. Unfortunately, high-quality labeled data usually require expensive manual work, and are therefore not readily available in many real scenarios. In this paper, we propose a novel learning paradigm for ER, called \emph{gradual machine learning}, which aims to enable effective machine labeling without the requirement for manual labeling effort. It begins with some easy instances in a task, which can be automatically labeled by the machine with high accuracy, and then gradually labels more challenging instances by iterative factor graph inference. In gradual machine learning, the hard instances in a task are gradually labeled in small stages based on the estimated evidential certainty provided by the labeled easier instances. Our extensive experiments on real data have shown that the performance of the proposed approach is considerably better than its unsupervised alternatives, and highly competitive compared to the state-of-the-art supervised techniques. Using ER as a test case, we demonstrate that gradual machine learning is a promising paradigm potentially applicable to other challenging classification tasks requiring extensive labeling effort.
\end{abstract}


\section{Introduction} \label{sec:introduction}

\begin{figure}[h]
\centering
\subfigure{$T_1$}{
\includegraphics[width=1.0\linewidth]{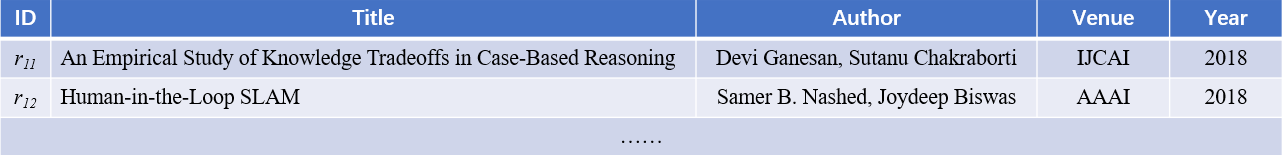}
}
\subfigure{$T_2$}{
\includegraphics[width=1.0\linewidth]{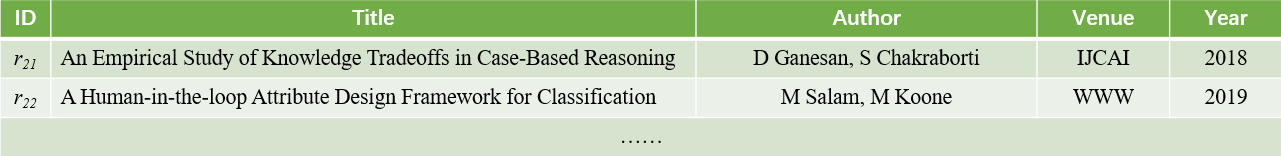}
}
\caption{An ER Example}
\label{fig:runningexample}
\end{figure}

  The task of entity resolution (ER) aims at finding the records that refer to the same real-world entity \citep{christen2012data}. Consider the running example shown in Figure~\ref{fig:runningexample}. ER needs to match the paper records between two tables, $T_1$ and $T_2$. The pair of $<r_{1i},r_{2j}>$, in which $r_{1i}$ and $r_{2j}$ denote a record in $T_1$ and $T_2$ respectively, is called an \emph{equivalent} pair if and only if $r_{1i}$ and $r_{2j}$ refer to the same paper; otherwise, it is called an \emph{inequivalent} pair. In the example, $r_{11}$ and $r_{21}$ are \emph{equivalent} while $r_{11}$ and $r_{22}$ are \emph{inequivalent}. The state-of-the-art solutions for ER were built on a variety of learning models (e.g. deep neural network (DNN)~\citep{mudgal18deep}), which require lots of accurately labeled training data. Unfortunately, high-quality labeled data usually require expensive manual work, and therefore, may not be readily available in many real scenarios.

\setlength{\textfloatsep}{8pt}
	
	It can be observed that the dependence of the existing supervised learning models on high-quality labeled data is not limited to
the task of ER. The dependence is actually crucial for their huge success in various domains (e.g. image and speech recognition~\citep{Yu2014Automatic}). However, it has been well recognized that in some real scenarios, where high-quality labeled data is scarce, their efficacy can be severely compromised. To address the limitation resulting from such dependence, we propose a novel learning paradigm, called \emph{gradual machine learning}, in which \emph{gradual} means proceeding in small stages. Gradual machine learning aims to enable effective machine labeling without the requirement for manual labeling effort. Inspired by the gradual nature of human learning, which is adept at solving the problems with increasing hardness, it begins with some easy instances in a task, which can be automatically labeled by the machine with high accuracy, and then gradually reasons about the labels of the more challenging instances based on the observations provided by the labeled easier instances.

  We note that there already exist many learning paradigms for a variety of classification tasks, including transfer learning \citep{Sinno2010a}, lifelong learning \citep{Chen2018Lifelong}, curriculum learning \citep{Bengio2009Curriculum} and self-training learning~\citep{selftraining} to name a few. Transfer learning focused on using the labeled training data in a domain to help learning in another target domain. Lifelong learning studied how to leverage the knowledge mined from past tasks for the current task. Curriculum learning investigated how to organize a curriculum (the presenting order of training examples) for better performance. Self-training learning aimed to improve the performance of a supervised learning algorithm by incorporating unlabeled data into the training data set. More recently, Snorkel \citep{Ratner2017SnorkelSigmod} aimed to enable automatic and massive machine labeling by specifying a wide variety of labeling functions. The results of machine labeling were supposed to be fed to DNN for model training. However, the following two properties of gradual machine learning make it fundamentally different from the existing learning paradigms:
\begin{itemize}	
\item Distribution misalignment between easy and hard instances in a task. Gradual machine learning processes the instances in the increasing order of hardness. Its scenario does not satisfy the i.i.d (independent and identically distributed) assumption underlying most existing machine learning models: the labeled easy instances are not representative of the unlabeled harder instances. The distribution misalignment between the labeled and unlabeled instances renders most existing learning models unfit for gradual machine learning.
\item Gradual learning by small stages in a task. Gradual machine learning proceeds in small stages. At each stage, it typically labels only one instance based on the evidential certainty provided by the labeled easier instances. The process of iterative labeling can be performed in an unsupervised manner without requiring any human intervention.
\end{itemize}

   We summarize the major contributions of this paper as follows:
\begin{enumerate}
  \item We propose a novel learning paradigm of Gradual Machine Learning (GML), which can effectively eliminate the requirement for manual labeling effort for the challenging classification tasks;
	\item We present a technical solution based on the proposed paradigm for entity resolution. We present a package of techniques, including easy instance labeling, feature extraction and influence modeling, and gradual inference, to enable effective gradual machine learning for ER.
  \item Our extensive experiments on real data have validated the efficacy of the proposed approach. Our empirical study has shown that the performance of the proposed approach is considerably better than the unsupervised alternatives, and highly competitive compared to the state-of-the-art supervised techniques. It also scales well with workload size.
\end{enumerate}

	Note that a prototype of the proposed GML solution for ER has been presented in the demo paper of~\citep{GMLDemo}. Besides providing with more technical details on the GML solution for ER, this technical paper makes the following new contributions:
\begin{enumerate}
  \item We propose a scalable solution for gradual inference. The solution consists of three steps, measurement of evidential support, approximate estimation of inference probability, and construction of inference subgraph. We have presented the detailed techniques for each of the three steps.
	\item We have evaluated the performance sensitivity of the proposed solution w.r.t various algorithmic parameters and its scalability. Our experimental results have shown that the proposed solution performs robustly w.r.t the parameters and it scales well with workload size.
\end{enumerate}	
	
   It is also noteworthy that we have recently applied the paradigm of gradual machine learning on the task of aspect-level sentiment analysis~\citep{GMLALSA}. Similar to the task of ER, the performance of GML has been shown to be highly competitive compared to the state-of-the-art DNN techniques.
	The rest of this paper is organized as follows: Section~\ref{sec:relatedwork} reviews related work. Section~\ref{sec:task} defines the task of ER. Section~\ref{sec:paradigm} introduces the general learning paradigm. Section~\ref{sec:solution} proposes the technical solution for ER. Section~\ref{sec:scalableER} presents the solution of scalable gradual inference for ER. Section~\ref{sec:experiments} presents our empirical evaluation results. Finally, Section~\ref{sec:conclusion} concludes this paper with some thoughts on future work.

\section{Related Work} \label{sec:relatedwork}

{\bf Machine Learning Paradigms}
\vspace{0.05in}

    Many machine learning paradigms have been proposed for a wide variety of classification tasks. Here, our intention is not to exhaustively review all the work. We instead review those closely related to our work and emphasize their difference from gradual machine learning.
	
   Traditional supervised machine learning algorithms make predictions on the future data using statistical models that are trained on previously collected labeled training data~\citep{christen2008automatic}. In many real scenarios, the labeled data may be too few to build a good classifier. Semi-supervised learning~\citep{Blum1998Combining,Joachims1999Transductive} addresses this problem by making use of a large amount of unlabeled data and a small amount of labeled data. Similarly, as an autonomous supervised learning approach, self-supervised learning~\citep{selfsupervised} usually extracts and uses the naturally available relevant context and embedded meta data as supervisory signals. Active learning~\citep{arasu2010active,bellare2012active} is another special case of supervised learning in which a learning algorithm is able to interactively query the user (or some other information source) to obtain the desired outputs at new data points. The main advantage of active learning over traditional supervised learning is that it usually requires less labeled data for model training. Nevertheless, the efficacy of the aforementioned learning paradigms depends on the i.i.d assumption. Therefore, they can not be applied to the scenario of gradual machine learning.

	
	Curriculum learning (CL)~\citep{Bengio2009Curriculum} and self-paced learning (SPL)~\citep{Kumar2010SPL} are to some extent similar to gradual machine learning in that they were also inspired by the learning principle underlying the cognitive process in humans, which generally starts with learning easier aspects of a task, and then gradually takes more complex examples into consideration. However, both of them depend on a curriculum, which is a sequence of training samples essentially corresponding to a list of samples ranked in ascending order of learning difficulty. A major disparity between them lies in the derivation of the curriculum. In CL, the curriculum is assumed to be given by an oracle beforehand, and remains fixed thereafter. In SPL, the curriculum is instead dynamically generated by the learner itself, according to what the learner has already learned. Online learning \citep{Kivinen2004Online} and incremental learning \citep{Schlimmer1986IL} have also been proposed for the scenarios where training data only becomes available gradually over time or its size is out of system memory limit. They were usually used to update the best predictor for future data at each step, as opposed to the batch learning techniques which generate the best predictor by learning on the entire training data set at once. It is worthy to point out that based on the traditional supervised learning models, all these learning paradigms depend on the i.i.d assumption and require good-coverage training examples for their efficacy. 
	
	In contrast, transfer learning~\citep{Sinno2010a}, allows the distributions of the data used in training and testing to be different. It focuses on using the labeled training data in a domain to help learning in another target domain. The other learning techniques closely related to transfer learning include lifelong learning~\citep{Chen2018Lifelong} and multi-task learning~\citep{Caruana1997Multitask}. Lifelong learning is similar to transfer learning in that it also focused on leveraging the experience gained on the past tasks for the current task. However, different from transfer learning, it usually assumes that the current task has good training data, and aims to further improve the learning using both the target domain training data and the knowledge gained in past learning. Multi-task learning instead tries to learn multiple tasks simultaneously even when they are different. A typical approach for multi-task learning is to uncover the pivot features shared among multiple tasks. However, all these learning paradigms can not be applied for the scenario of gradual machine learning. Firstly, focusing on unsupervised learning within a task, gradual machine learning does not enjoy the access to good labeled training data or a well-trained classifier to kick-start learning. Secondly, the existing techniques transfer instances or knowledge between tasks in a batch manner; they do not support gradual learning by small stages on the instances with increasing hardness within a task.


	
\vspace{0.05in}
\hspace{-0.22in}{\bf Work on Entity Resolution}
\vspace{0.05in}

     Research effort on unsupervised entity resolution were mainly dedicated to devising various distance functions to measure pair-wise similarity~\citep{Monge96thefield}. However, it has been empirically shown that the efficacy of these unsupervised techniques is limited~\citep{Bilenko2003ANM}. Alternatively, the supervised techniques viewed the problem of ER as a binary classification task and then applied various learning models (e.g. SVM \citep{arasu2010active,bellare2012active}, native Bayesian~\citep{berger1985statistical}, and DNN models~\citep{mudgal18deep}) for the task. Compared with the unsupervised alternatives, they can effectively improve the quality of entity resolution to some extent. However, good performance of the supervised techniques depends on the presence of a large quantity of accurately labeled training data, which may not be readily available in real applications.
	
	 The progressive paradigm for ER~\citep{altowim2014progressive,whang2013pay} has also been proposed for the application scenarios in which ER should be processed efficiently but does not necessarily require to generate high-quality results. Taking a pay-as-you-go approach, it studied how to maximize result quality given a pre-specified resolution budget. It fulfilled the purpose by constructing various resolution hints that
can be used by a variety of existing ER algorithms as a guidance for which entities to resolve first. It is worthy to point out that the target scenario of progressive ER is different from that of gradual machine learning, whose major challenge is to label the instances with increasing hardness without resolution budget.

     It has been well recognized that pure machine algorithms may not be able to produce satisfactory results in practical scenarios~\citep{li2016crowdsourced}. Therefore, many researchers ~\citep{chai2016cost,Chu2015Katara,Firmani2016Online,Getoor2012Entity,gokhale2014corleone,mozafari2014scaling,verroios2017waldo,vesdapunt2014crowdsourcing,wang2012crowder,wang2015crowd,whang2013question,YangJuFan2018Cost} have studied how to crowdsource an ER workload. While these researchers addressed the challenges specific to crowdsourcing, we instead investigate a different problem in this paper: how to enable unsupervised gradual machine learning.

\section{Task Statement}  \label{sec:task}

   Entity resolution reasons about the equivalence between two records. Two records are deemed to be equivalent if and only if they correspond to the same real-world entity; otherwise, they are deemed to be inequivalent. We call a record pair an equivalent pair if and only if its two records are equivalent; otherwise, it is called an inequivalent pair. Given an ER workload consisting of record pairs, a solution labels each pair in the workload as {\em matching} or {\em unmatching}.
	
\begin{table}[!h]
\centering
\caption{Frequently Used Notations.}
\label{tb:notations}
\begin{tabular}{|l|l|}
\hline
 Notation   & Description \\ \hline
 $D$        & an ER workload consisting of record pairs \\ \hline
 $D_i$      & a subset of $D$ \\ \hline
 $S$, $S_i$ & a labeling solution for $D$ \\ \hline
 $d$, $d_i$ & a record pair in $D$ \\ \hline
 $TN(D_i)$  & the total number of pairs in $D_i$ \\ \hline
 $EN(D_i)$  & the total number of equivalent pairs in $D_i$ \\ \hline
 $P(d_i)$   & the estimated equivalence probability of $d_i$ \\ \hline
 $f$, $f_i$ & a feature of record pair \\ \hline
 $F$, $F_i$ & a feature set \\ \hline
 $D_f$      & the set of record pairs having the feature $f$ \\ \hline
\end{tabular}
\end{table}

    For the sake of presentation simplicity, we summarize the frequently used notations in Table.~\ref{tb:notations}. As usual, we measure the quality of a labeling solution by the metrics of precision and recall. The metric of precision denotes the fraction of equivalent pairs among all the pairs labeled as {\em matching}, while recall denotes the fraction of the equivalent pairs labeled as \emph{matching} among all the equivalent pairs. Given an ER workload $D$ and a labeling solution $S$, suppose that $D_+$ denotes the set of record pairs labeled as \emph{matching}, and $D_-$ denotes the set of record pairs labeled as \emph{unmatching}. Then, the achieved precision level of $S$ on $D$ can be represented by
\begin{equation} \label{eq:precision}
    \emph{precision(D,S)} = \frac{EN(D_+)}{TN(D_+)},
\end{equation}
in which $TN(\cdot)$ denotes the total number of pairs in a set, and $EN(\cdot)$ denotes the total number of equivalent pairs in a set. Similarly, the achieved recall level of $S$ can be represented by
\begin{equation} \label{eq:recall}
    \emph{recall(D,S)} = \frac{EN(D_+)}{EN(D_+) + EN(D_-)}.
\end{equation}
The overall quality of entity resolution is usually measured by the unified metric of F-1 as represented by
\begin{equation}
  f_1(D,S)=\frac{2}{\frac{1}{precision(D,S)} + \frac{1}{recall(D,S)}}.
\end{equation}

  Finally, the task of entity resolution is defined as follows:

\begin{definition}
\label{problemsetting}
{\bf [Entity Resolution].}  Given a workload consisting of record pairs, $D=\{d_1, d_2, \cdots, d_n\}$, the task of entity resolution is to give a labeling solution $S$ for $D$ such that $f_1(D,S)$ is maximized.
\end{definition}


\section{Learning Paradigm}  \label{sec:paradigm}

\setlength{\textfloatsep}{8pt}
\begin{figure}[ht]
\centering
\includegraphics[width=0.99\linewidth]{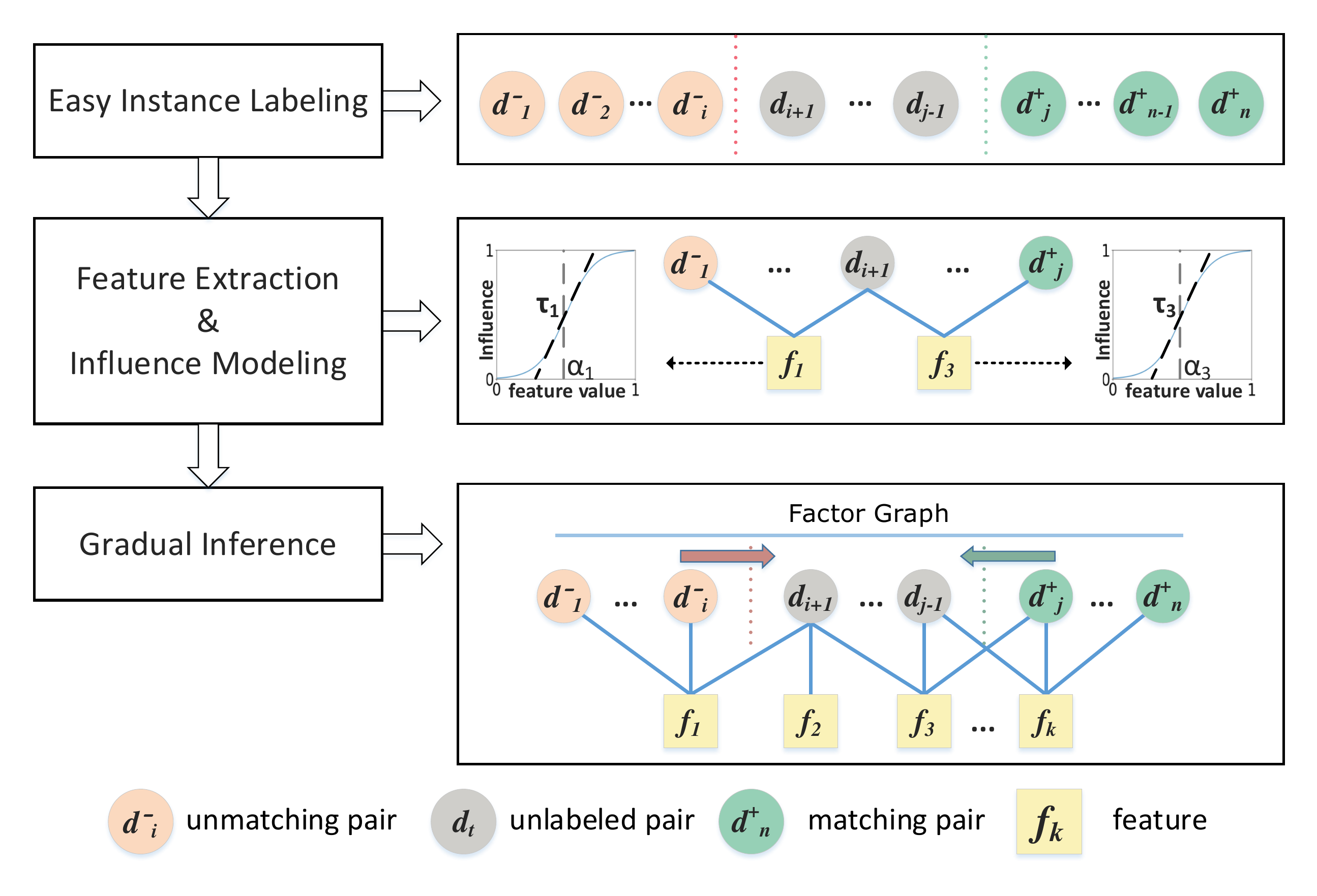}
\caption{Paradigm Overview.}
\label{fig:framework}
\end{figure}

   The process of gradual machine learning, as shown in Figure~\ref{fig:framework}, consists of the following three essential steps:

\begin{itemize}
\item {\bf Easy Instance Labeling.} Given a classification task, it is usually very challenging to accurately label all the instances in the task without good-coverage training examples. However, the work can become much easier if we only need to automatically label some easy instances in the task. In the case of ER, while the pairs with the medium similarities are usually challenging for machine labeling, highly similar (resp. dissimilar) pairs have fairly high probabilities to be equivalent (resp. inequivalent). They can therefore be chosen as easy instances. In real scenarios, easy instance labeling can be performed based on the simple user-specified rules or the existing unsupervised learning techniques. For instance, in unsupervised clustering, an instance close to the center of a cluster in the feature space can be considered an easy instance, because it only has a remote chance to be misclassified. Gradual machine learning begins with the observations provided by the labels of easy instances. Therefore, the high accuracy of automatic machine labeling on easy instances is critical for its ultimate performance on a given task.

\item {\bf Feature Extraction and Influence Modeling.} Features serve as the medium to convey the knowledge obtained from the labeled easy instances to the unlabeled harder ones. This step extracts the common features shared by the labeled and unlabeled instances. To facilitate effective knowledge conveyance, it is desirable that a wide variety of features are extracted to capture as much information as possible. For each extracted feature, this step also needs to model its influence over the labels of its relevant instances.

 \item {\bf Gradual Inference.} This step gradually labels the instances with increasing hardness in a task. Since the scenario of gradual learning does not satisfy the i.i.d assumption, we propose to fulfill gradual learning from the perspective of evidential certainty. As shown in Figure~\ref{fig:framework}, we construct a factor graph, which consisting of the labeled and unlabeled instances and their common features. Gradual learning is conducted over the factor graph by iterative factor graph inference. At each iteration, it chooses the unlabeled instance with the highest degree of evidential certainty for labeling. The iteration is repeatedly invoked until all the instances in a task are labeled. Note that in gradual inference, a newly labeled instance at the current iteration would serve as an evidence observation in the following iterations.

\end{itemize}

  The framework laid out in Figure~\ref{fig:framework} is general in that additional technical work is required for building a practical solution for a classification task. It is noteworthy that many techniques proposed in the existing learning models can be potentially leveraged in the different steps of gradual machine learning. For instance, the existing rule-based and unsupervised clustering techniques can be used to label easy instances. There also exist many techniques to extract the features for supervised and unsupervised learning. They can be potentially used in the step of feature extraction and influence modeling.
	
	In the rest of this paper, we will present the technical solution for each of the three steps of GML with ER as the target task. As we have shown in~\citet{GMLALSA}, the paradigm can also be applicable to other classification tasks, but the details of its technical solutions may be different.

\section{Solution for ER} \label{sec:solution}

\subsection{Easy Instance Labeling}

   Given an ER task consisting of record pairs, the solution identifies the easy instances by the simple rules specified on record similarity.
The set of easy instances labeled as \emph{matching} is generated by setting a high lowerbound on record similarity. Similarly, the set of easy instances labeled as \emph{unmatching} is generated by setting a low upperbound on record similarity. To explain the effectiveness of the rule-based approach, we introduce the monotonicity assumption of precision, which was first defined in~\citet{arasu2010active}, as follows:
	
\begin{assumption}[Monotonicity of Precision]
  A value interval $I_i$ is dominated by another interval $I_j$, denoted by $I_i\preceq I_j$, if every value in $I_i$ is less than every value in $I_j$. We say that precision is monotonic with respect to a pair metric if for any two value intervals $I_i\preceq I_j$ in [0,1], we have $P(I_i)\leq P(I_j)$, in which $P(I_i)$ denotes the equivalence precision of the set of instance pairs whose metric values are located in $I_i$.
\label{monotonicity}
\end{assumption}

  With the metric of pair similarity, the underlying intuition of Assumption~\ref{monotonicity} is that the more similar two records are, the more likely they refer to the same real-world entity. According to the monotonicity assumption, we can \emph{statistically} state that a pair with a high (resp. low) similarity has a correspondingly high probability of being an equivalent (resp. inequivalent) pair. These record pairs can be deemed to be easy in that they can be automatically labeled by the machine with high accuracy. In comparison, the instance pairs having the medium similarities are more challenging because labeling them either way by the machine would introduce considerable errors.
	
\begin{figure}[h!]
\centering
\includegraphics[width=0.4\linewidth]{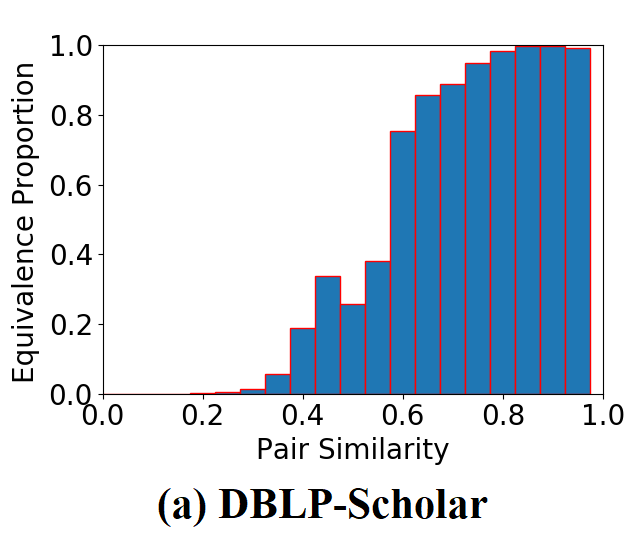}
\includegraphics[width=0.4\linewidth]{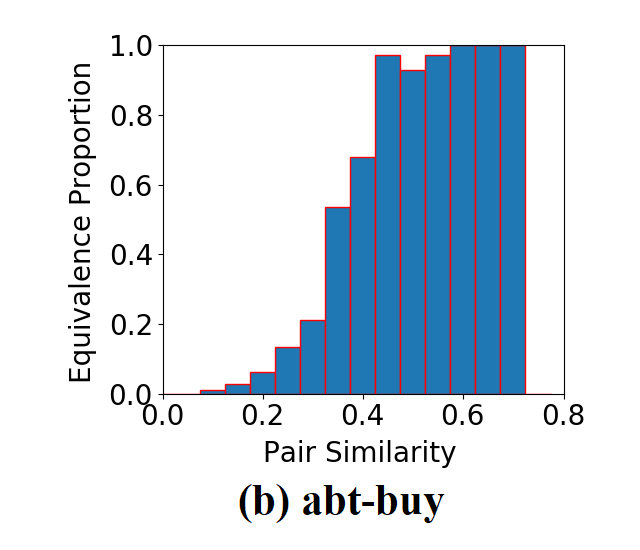}
\caption{Empirical Validation of the Monotonicity Assumption.}
\label{fig:monotonicity}
\end{figure}

	We have empirically validated the monotonicity assumption on the real datasets of DBLP-Scholor\footnote{available at https://dbs.uni-leipzig.de/file/DBLP-Scholar.zip} and Abt-Buy\footnote{available at https://dbs.uni-leipzig.de/file/Abt-Buy.zip}. The precision levels of different similarity intervals are shown in Figure~\ref{fig:monotonicity}. It can be observed that statistically speaking, precision increases with similarity value with notably rare exceptions. The proposed approach assumes that the monotonicity of precision is a statistical trend. It however does \emph{not} expect that the monotonicity assumption can be \emph{strictly} satisfied on real data. On DBLP-Scholar, if the similarity lowerbound is set to be 0.8, the achieved precision is 0.992, nearly 100\%. On the other hand, if the similarity upperbound is set to be 0.3, the ratio of \emph{inequivalent} pairs is similarly very high at 0.997. We have the similar observation on Abt-Buy.
	
	It is noteworthy that given a machine metric for a classification task, the monotonicity assumption of precision actually underlies its effectiveness as a classification metric. Therefore, the easy instances in an ER task can be similarly identified by other classification metrics. However, for illustrative purpose, we use pair similarity as the example of machine metric in this paper.

\subsection{Feature Extraction and Influence Modeling}	

   The guiding principle of feature extraction is to extract a wide variety of discriminating features that can capture as much information as possible from the record pairs. Given an ER workload, we extract the following two types of features from its pairs:
\begin{enumerate}	
\item Attribute value similarity. This type of feature measures a pair's value similarity at each record attribute. Different attributes may require different similarity metrics. For instance, on the DBLP-Scholar dataset, the appropriate metric for the \emph{venue} attribute is the edit distance, while the appropriate metric for the \emph{title} attribute is instead a hybrid metric combining Jacard similarity and edit distance. For long string attributes (e.g., the \emph{title} attribute in the literature records and the attribute of \emph{product description} in the product records), we also measure the the Longest Common Substring (LCS) similarity between two records, which refer to the number of tokens in the longest common substring. Given a record pair with long string value attributes, its length of LCS can usually, to a large extent, affect its equivalence probability. 

\item The tokens occurring in both records or in one and only one record. Suppose that we denote a token by $o_i$, the feature that $o_i$ occurs in both records by $Same(o_i)$, and the feature that $o_i$ occurs in one and only one record by $Diff(o_i)$. Note that the feature of $Same(o_i)$ serves as evidence for equivalence, while the feature of $Diff(o_i)$ indicates the opposite. Unlike the previous two types of features, which treat attribute values as a whole, this type of feature considers the influence of each individual token on pair equivalence probability. Since not every token is highly discriminating (or indicative of entity identity), we filter the tokens in a workload by the metric of IDF (inverse document frequency).
\end{enumerate}

\begin{figure}[ht]
\centering
\includegraphics[width=0.8\linewidth]{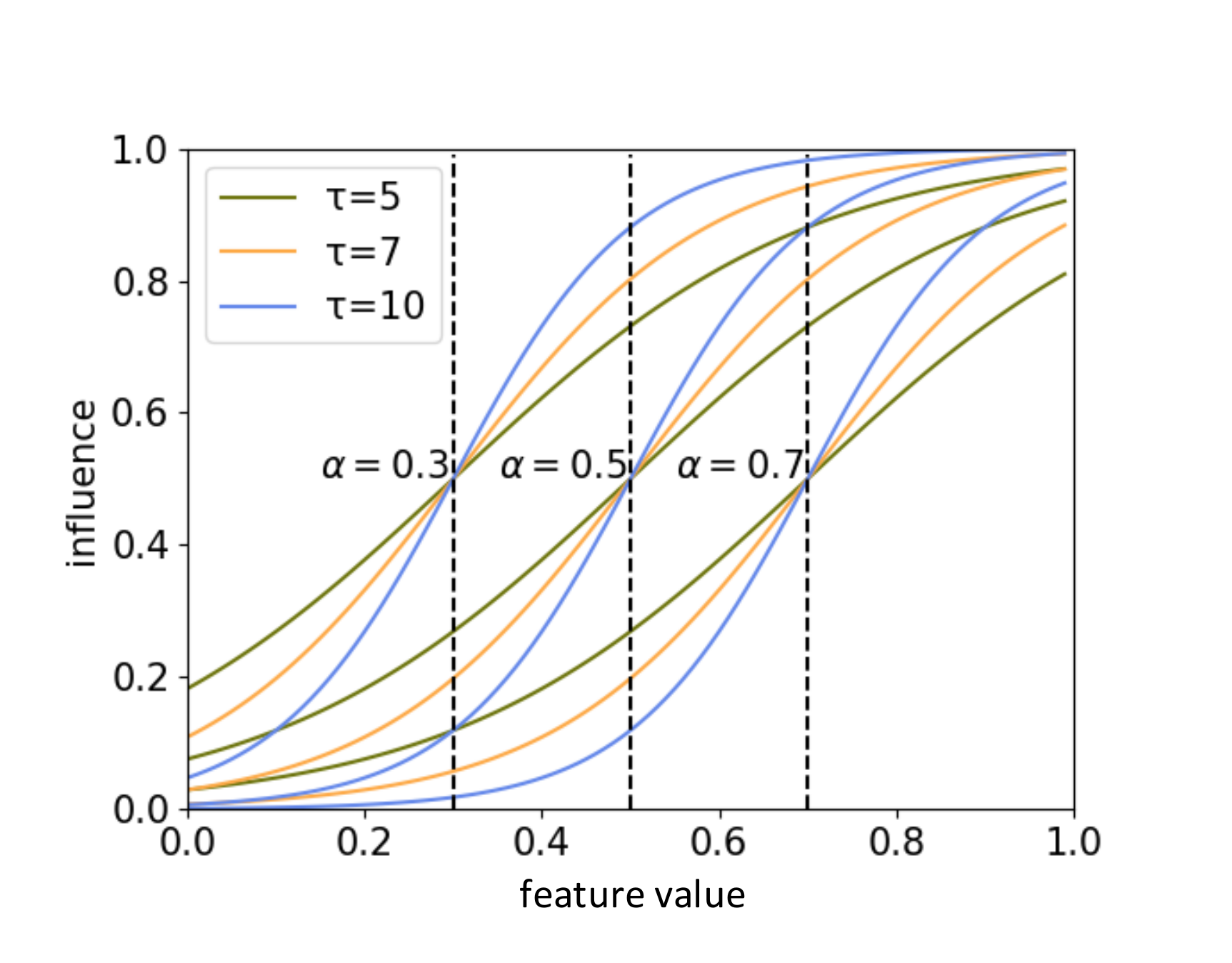}
\caption{the Examples of Sigmoid Function.}
\label{fig:sigmoid}
\end{figure}

  The aforementioned two types of features can provide good coverage of the information contained in record pairs. We observe that both of them can be supposed to satisfy the monotonicity assumption of precision.  Therefore, for each feature, we model its influence over pair labels by a monotonous sigmoid function with two parameters, $\alpha$ and $\tau$ as shown in Figure~\ref{fig:sigmoid}, which denote the $x$-value of the function's midpoint and the steepness of the curve respectively. The $x$-value of the sigmoid function represents the value of a pair w.r.t the corresponding feature, and the $y$-value represents the equivalent probability of a pair as estimated by the feature. Formally, given a feature $f$ and a pair $d$, the influence of $f$ w.r.t $d$ is represented by
\begin{equation}
\label{eq:sigmoid}
  P_f(d) = \frac{1}{{1 + {e^{ - {\tau _f}(x_f(d) - {\alpha _f})}}}},
\end{equation}
in which $x_f(d)$ represents $f$'s value w.r.t $d$. Since the second type of features has the constant value of 1, we first align them with record similarity and then model their influence by sigmoid functions.

	  We illustrate the sigmoid function by the examples shown in Figure~\ref{fig:sigmoid}. It can be observed that different value combinations of $\alpha$ and $\tau$ can result in vastly different influence curves. Given a sigmoid model, gradual machine learning essentially reasons about the labels of the middle points, which correspond to the hard instances, provided with the labels of the more extreme points at both sides, which correspond to the easy instances. If it were not for the monotonicity assumption, estimating the labels of the middle points by regression would be too erroneous because the more extreme observations at both sides are not their valid representatives. Our solution overcomes this hurdle by assuming monotonicity of precision and proceeding in small stages, in each of which the regression results of only a few instances close to the labeled easy instances are considered for equivalence reasoning. Fortunately, monotonicity of precision is a universal assumption underlying the effectiveness of the existing machine metrics for classification tasks. Therefore, our proposed solution for modeling feature influence can be potentially generalized for other classification tasks.

\subsection{Gradual Inference}
\begin{figure}[ht]
\centering
\includegraphics[width=0.9\linewidth]{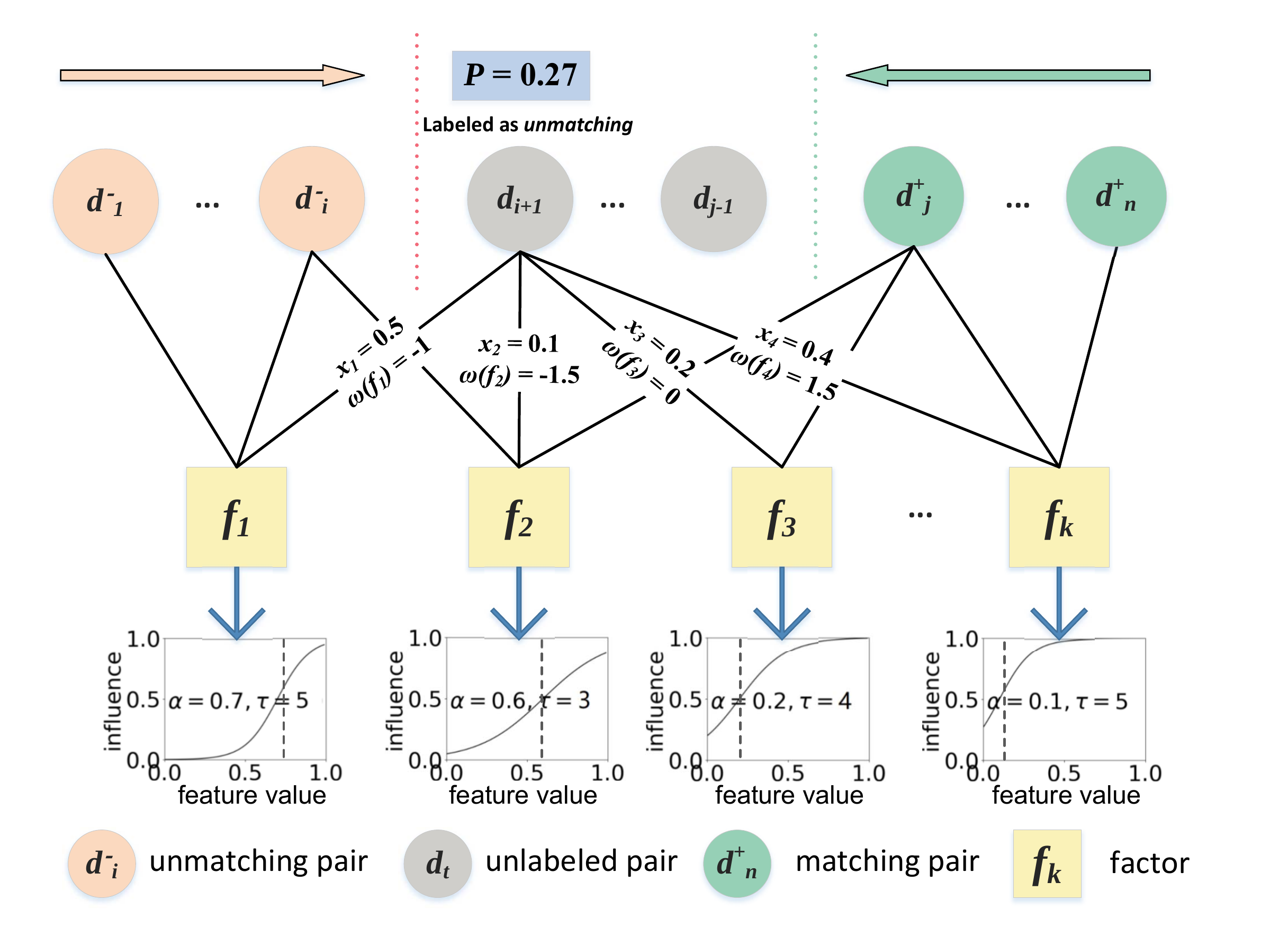}
\caption{An Example of Factor Graph.}
\label{fig:factorgraph}
\end{figure}

    To enable gradual machine learning, we construct a factor graph, $G$, which consists of the labeled easy instances, the unlabeled hard instances and their common features. Gradual machine learning is attained by iterative factor graph inference on $G$. In $G$, the labeled easy instances are represented by the \emph{evidence variables}, the unlabeled hard instances by the \emph{inference variables}, and the features by the \emph{factors}. The value of each variable represents its corresponding pair's equivalence probability. An evidence variable has the constant value of 0 or 1, which indicate the status of \emph{unmatching} and \emph{matching} respectively. It participates in gradual inference, but its value remains unchanged during the inference process. The values of the inference variables should instead be inferred based on $G$. 

  	An example of factor graph is shown in Figure~\ref{fig:factorgraph}. Each variable has multiple factors, each of which corresponds to a feature. Since a feature can be shared among multiple pairs, for presentation simplicity, we represent a feature by a single factor and connect it to multiple variables. Note that given a feature $f$ and a pair $d$, the influence of $f$ w.r.t $d$ is represented by the sigmoid function of
\begin{equation}
  P_f(d) = \frac{1}{{1 + {e^{ - {\tau _f}(x_f(d) - {\alpha _f})}}}},
\end{equation}
in which $x_f(d)$ represents $f$'s value w.r.t $d$, which is known beforehand, and $\tau _f$ and $\alpha _f$ represents the parameters of a sigmoid function, which need to be learned. Accordingly, in the factor graph, we represent the factor weigh of $f$ w.r.t $d$ by
\begin{equation}
\label{eq:factorweight}
  {\omega _f}(d) = {\theta _f}(d)\cdot \log (\frac{{{P_f}(d)}}{{1 - {P_f}(d)}}) = {\theta _f}(d)\cdot {\tau _f}({x_f}(d) - {\alpha _f}),
\end{equation}
in which $\log (\cdot)$ codes the estimated influence of $f$ on $d$ by sigmoid regression, and ${\theta _f}(d)$ represents the confidence on influence estimation. In practical implementation, we can estimate ${\theta _f}(d)$ based on the theory of regression error bound~\citep{Chen1994Empirical}. More details on the estimation of ${\theta _f}(d)$ will be discussed in Subsection~\ref{sec:evidentialsupport}.

Denoting the feature set of a pair $d$ by $F_d$, a factor graph infers the equivalence probability of $d$, $P(d)$, by:
\begin{equation}
P(d) = \frac{{\prod\limits_{f \in {F_d}} {{e^{\omega _f(d)}}} }}{{1 + \prod\limits_{f \in {F_d}} {{e^{\omega _f(d)}}} }}.
\end{equation}

  The process of gradual inference essentially learns the parameter values ($\alpha$ and $\tau$) of all the features such that the inferred results maximally match the evidence observations on the labeled instances. Formally, the objective function can be represented by
\begin{equation}
(\hat \alpha ,\hat \tau ) = \arg \mathop {\min }\limits_{\alpha ,\tau }  - \log \sum\limits_{{V_I}} {{P_{\alpha ,\tau }}(\Lambda ,{V_I})} ,
\end{equation}
in which $\Lambda$ denotes the observed labels of evidence variables, $V_I$ denotes the inference variables in $G$, and ${{P_{\alpha ,\tau }}(\Lambda ,{V_I})}$ denotes the joint probability of the variables in $G$. Since the variables in $G$ are conditionally independent, ${P_{\alpha ,\tau }}(\Lambda ,{V_I})$ can be represented by:
\begin{equation}
\label{eq:independent}
{P_{\alpha ,\tau }}(\Lambda ,{V_I}) = \mathop \prod \limits_{d \in \Lambda \cup {V_I}} {P_{\alpha ,\tau }}(d).
\end{equation}
Accordingly, the objective function can be simplified into
\begin{equation}
\label{eq:independentmle}
(\hat \alpha ,\hat \tau ) = \arg \mathop {\min }\limits_{\alpha ,\tau }  - \sum\limits_{d \in \Lambda } \log {{P_{\alpha ,\tau }}(d)}.
\end{equation}
Considering the inequality between the observations of two classes, we also weight the observations of two classes to perform the weighted maximum likelihood estimation as in~\citet{weightedMLE}. Specifically, given the factor graph consisting of $n_-$ unmatching and $n_+$ matching observations, the weights of the unmatching and matching observations are set to be $1$ and $\frac{n_-}{n_+}$ respectively. Finally, the objective function can be represented by
\begin{equation}
\label{eq:weightmle}
(\hat \alpha ,\hat \tau ) = \arg \mathop {\min }\limits_{\alpha ,\tau }  - \sum\limits_{d \in \Lambda } t_d \cdot \log {{P_{\alpha ,\tau }}(d)},
\end{equation}
in which $t_d=1$ if $d$ is labeled as unmatching, and $t_d=\frac{n_-}{n_+}$ if $d$ is labeled as matching.

	Given a factor graph, $G$, at each stage, gradual inference first reasons about the parameter values of the features and the equivalence probabilities of the unlabeled pairs by maximum likelihood, and then labels the unlabeled pair with the highest degree of evidential certainty. We define evidential certainty as the inverse of entropy~\citep{Shannon}, which is formally defined by
\begin{equation}
  H(d) =  - (P(d) \cdot {{\log }_2}P(d) + (1 - P(d)) \cdot {{\log }_2}(1 - P(d))),
\end{equation}
in which $H(d)$ denotes the entropy of $d$. According to the definition, the degree of evidential certainty varies inversely with the estimated value of entropy. The value of $H(d)$ reaches its maximum when $P(d)=0.5$, and it decreases as the value of $P(d)$ becomes more extreme (close to 0 or 1). Therefore, at each iteration, gradual inference selects the instance pair with the minimal entropy for labeling. It labels the chosen instance pair as \emph{matching} if $P(d)\ge 0.5$, and as \emph{unmatching} if $P(d)<0.5$. An inference variable once labeled would become an evidence variable and serve as an evidence observation in the following iterations. The iteration is repeatedly invoked until all the inference variables are labeled. Since each iteration of GML would label only one instance, therefore the total number of iterations increases linearly with workload size.

    In our implementation, we used the platform of \emph{SciPy} \citep{scipy} to implement the parameter optimization process. The process searches to identify the optimal parameters which can minimize the objective function as shown in Eq.~\ref{eq:weightmle}.
    To avoid overfitting, the search range of the midpoint parameter $\alpha _f$ of a feature $f$ is set to be between the expectations of the feature values of the pairs labeled as \emph{unmatching} and \emph{matching}. The search space of the parameter $\tau _f$ is set to be $[0, 10]$ for all the features.

  Unfortunately, repeated inference by maximum likelihood estimation over a large-sized factor graph of the whole variables is usually very time-consuming~\citep{Zhou2016ArchimedesOne}. As a result, the above-mentioned approach can not scale well with a large ER workload. In the next section, we will propose a scalable approach that can effectively fulfill gradual learning without repeatedly inferring over the entire factor graph.

\section{Scalable Gradual Inference}	\label{sec:scalableER}

   The scalable solution is crafted based on the following observations:
\begin{itemize}
   \item Many unlabeled inference variables in the factor graph may be only weakly linked through the factors to the evidence variables. Due to lack of evidential support, their inferred probabilities would be quite ambiguous, i.e. close to 0.5. As a result, at each stage, only the inference variables that have received considerable support from the evidence variables need to be considered for labeling;
   \item With regard to the probability inference of a single variable $v$ in a large factor graph, it can be effectively approximated by considering the potentially much smaller subgraph consisting of $v$ and its neighboring variables. The inference over the subgraph can usually be much more efficient than over the original entire graph.
\end{itemize}

\begin{algorithm}[t]
\caption{Scalable Gradual Inference}
\label{alg:gradualinference}
\While{there exists any unlabeled variable in $G$}
  {
    $V' \leftarrow$ all the unlabeled variables in $G$\;
		\For{$v\in V'$}
		{
		    Measure the evidential support of $v$ in $G$\;
		}
		Select top-m unlabeled variables with the most evidential support (denoted by $V_m$) \;
		\For{$v\in V_m$}
		{
		    Estimate the probability of $v$ in $G$ by approximation\;
		}
		Select top-$k$ certain variables in terms of entropy in $V_m$ based on the approximate probabilities (denoted by $V_k$) \;
		\For{$v\in V_k$}
		{
			Compute the probability of $v$ in $G$ by the factor graph inference over a subgraph of $G$\;
		}
		Label the variable with the minimal entropy in $V_k$\;
  }
\end{algorithm}
		
	 The process of scalable gradual inference is sketched in Algorithm~\ref{alg:gradualinference}. It first selects the top-$m$ unlabeled variables with the most evidential support in $G$ as the candidates for probability inference. To reduce the invocation of maximum likelihood estimation, it then approximates probability inference by an efficient algorithm on the $m$ candidates. Finally, it infers via maximum likelihood the probabilities of only the top-$k$ most promising unlabeled variables among the $m$ candidates. For each variable in the final set of $k$ candidates, its probability is not inferred over the entire graph of $G$, but over a potentially much smaller subgraph.

	The rest of this section is organized as follows: Subsection~\ref{sec:evidentialsupport} presents the technique to measure evidential support. Subsection~\ref{sec:approximateinference} presents the approximation algorithm to efficiently rank the inference probabilities of unlabeled variables. Subsection~\ref{sec:subgraphconstruction} describes how to construct an inference subgraph for a target unlabeled variable.

\subsection{Measurement of Evidential Support} \label{sec:evidentialsupport}

  Since the influence of a feature over the pairs is modeled by a sigmoid function, we consider the evidential support that an unlabeled variable receives from a feature as the confidence on the regression result provided by its corresponding function, denoted by ${\theta _f}(d)$. Note that ${\theta _f}(d)$ is also used to compute the confidence-aware factor weight in Eq.~\ref{eq:factorweight}. Given an unlabeled variable, $d$, we first estimate its evidential support provided by each of its factors based on the theory of regression error bound~\citep{Chen1994Empirical}, and then aggregate them to estimate its overall evidential support based on the Dempster-Shafer theory~\citep{Shafer1976DST}.

	Formally, for the influence estimation of a single feature $f$ on the variables, the process of parameter optimization corresponds to a linear regression between the natural logarithmic coded influence in Eq.~\ref{eq:factorweight}, hereinafter denoted by ${l_f}(d)$, and the feature value $x_f(d)$, as follows
\begin{equation}
\label{eq:regression}
{l_f}(d) = {\tau _f} \cdot {x_f}(d) - {\tau _f} \cdot {\alpha _f} + \varepsilon ,
\end{equation}
in which $\varepsilon$ denotes the regression residual. The parameters $\alpha_f$ and $\tau_f$ are optimized by minimizing the regression residual as follows:
\begin{equation}
\label{eq:regressionparameter}
({\hat \alpha _f},{\hat \tau _f}) = \arg \mathop {\min }\limits_{{\alpha _f},{\tau _f}} \sum\limits_{d \in {\Lambda _f}} {t_d \cdot {{({l_f}(d) - ({\tau _f} \cdot {x_f}(d) - {\tau _f} \cdot {\alpha _f}))}^2}} ,
\end{equation}
in which $\Lambda _f$ denotes the set of labeled pairs having the feature $f$. As in Eq.~\ref{eq:weightmle}, $t_d$ denotes the weights of matching and unmatching observations.


	 According to the theory of linear regression error bound, given a pair $d$, its prediction error bound $\delta ({l_f}(d))$ and the confidence level ${\theta _f}(d)$ satisfy the following formula
\begin{equation}
\label{eq:errorbound}
\begin{split}
&\delta ({l_f}(d)) =\\
&{t_{(1 - {\theta _f}(d))/2}}(|{\Lambda _f}| - 2) \cdot {\hat \sigma ^2} \cdot \sqrt {1 + \frac{1}{n} + \frac{{({x_f}(d) - {{\bar x}_f})}}{{\sum\limits_{{d_i} \in {\Lambda _f}} {({x_f}(} {d_i}) - {{\bar x}_f})}}} ,
\end{split}
\end{equation}
in which ${t_{(1 - {\theta _f}(d) )/2}}(|\Lambda _f| - 2)$ represents the Student's $t$-value with $|\Lambda _f| - 2$ degree of freedom at $(1 - {\theta _f}(d) )/2$ quantile, and
\begin{equation}
{\hat \sigma ^2} = \frac{1}{{|{\Lambda _f}| - 2}}\sum\limits_{{d_i} \in {\Lambda _f}} {{{({l_f}({d_i}) - ({{\hat \tau }_f} \cdot {x_f}({d_i}) - {{\hat \tau }_f} \cdot {{\hat \alpha }_f}))}^2}} ,
\end{equation}
and
\begin{equation}
{{\bar x}_f} = \frac{1}{{|\Lambda _f|}}\mathop \sum \limits_{d_i \in \Lambda _f} {x_f}(d_i).
\end{equation}

  Given an error bound of $\delta ({l_f}(d))$, we measure the evidential support of an unlabeled variable $d$ provided by $f$ by estimating its corresponding regression confidence level ${\theta _f}(d)$ according to Eq.~\ref{eq:errorbound}. Then, we use the classical theory of evidence, the Dempster-Shafer (D-S) theory~\citep{Shafer1976DST}, to combine the evidential support from different features and arrive at a degree of belief that takes into account all the available evidence. In our case, given a variable $v$, the evidential support provided by a feature $f$ can be considered to be the extent that $f$ supports the inference on the probability of $v$: a value of 1 means complete support while a value of 0 corresponds to the lack of any support. Suppose that an unlabeled variable $v$ has $l$ features, \{$f_1$,$\cdots$,$f_l$\}, and the evidential support $v$ receiving from $f_i$ is denoted by $\theta_i$. We first normalize the values of $\theta_i$ by $\frac{1+\theta_i}{2}$ so that $\theta_i$ falls into the value range of $[0.5,1]$. Then, by the Dempster's rule, we combine the evidential support of $v$ provided by its features by
\begin{equation}
\theta_v=\frac{\mathop\prod\limits_{1\leq i\leq l}{\theta_i}}{\mathop\prod\limits_{1\leq i\leq l}{\theta_i} + \mathop\prod\limits_{1\leq i\leq l}{(1-\theta_i)}}.
\end{equation}

  On time complexity, each iteration of evidential support measurement takes $O(n\cdot n_f)$ time, in which $n$ denotes the total number of instances in a task, and $n_f$ denotes the total number of extracted features. Therefore, we have Lemma~\ref{evidencecomplexity}, whose proof is straightforward, thus omitted here.

\begin{lemma} \label{evidencecomplexity}
    Given an ER task, the total computational cost of evidential support measurement can be represented by $O(n^2\cdot n_f)$.
\end{lemma}

\subsection{Approximate Estimation of Inferred Probability}	\label{sec:approximateinference}

    Due to the prohibitive cost of factor graph inference, at each iteration, reasoning about the probabilities of all the top-m inference variables ranked by evidential support via the factor graph inference may still be too time-consuming. Therefore, there is a need to efficiently approximate the inferred probabilities of these top-m variables such that only a small portion (top-k) of them needs to be inferred using factor graph inference.

As previously mentioned, the feature's natural logarithmic influence w.r.t a pair can be estimated by the linear regression value based on Eq.~\ref{eq:regression}. Therefore, we approximate the factor weight of $f$ w.r.t $d$, ${{\hat \omega }_f}(d)$, by
\begin{equation}
{{\hat \omega }_f}(d) = {\theta _f}(d) \cdot {{\hat \tau }_f}({x_f}(d) - {{\hat \alpha }_f}),
\end{equation}
in which ${\theta_f}(d)$ represents $f$'s normalized confidence level on the regression result w.r.t $d$ and ${\hat \tau }_f$, and ${\hat \alpha }_f$ are
the regression parameter values estimated by Eq.~\ref{eq:regressionparameter}. Accordingly, a pair's equivalence probability can be approximated by leveraging the approximate factor weights of all its features as follows
\begin{equation}
\hat P(d) = \frac{{\prod\limits_{f \in {F_d}} {{e^{{{\hat \omega }_f}(d)}}} }}{{1 + \prod\limits_{f \in {F_d}} {{e^{{{\hat \omega }_f}(d)}}} }},
\end{equation} 	
in which $F_d$ denotes the feature set of $d$.

  On time complexity, each iteration of approximate probability estimation takes $O(n_f \cdot m)$ time, in which $n_f$ denotes the total number of extracted features. Therefore, we have Lemma~\ref{approximatecomplexity}, whose proof is straightforward, thus omitted here.

\begin{lemma} \label{approximatecomplexity}
  Given an ER task, the total computational cost of approximate probability estimation can be represented by $O(n \cdot n_f \cdot m)$.
\end{lemma}				

  In practical implementation, due to the high efficiency of evidential support measurement and inference probability approximation, the number of candidate inference variables selected for approximate probability estimation ($m$) can be usually set to a large value (in the order of thousands). However, the number of candidate inference variables chosen for factor graph inference ($k$) is usually set to a much smaller value (in the order of tens), due to the inefficiency of factor graph inference. Our empirical evaluation in Section~\ref{sec:experiments} has showed that to a large extent, the performance of scalable gradual inference is not sensitive to the parameter settings of $m$ and $k$.
	
\subsection{Construction of Inference Subgraph} \label{sec:subgraphconstruction}

Given a target inference variable $v_i$ in a large factor graph $G$, inferring $v_i$'s equivalence probability over the entire graph is usually very time-consuming. Fortunately, it has been shown that factor graph inference can be effectively approximated by considering the subgraph consisting of $v_i$ and its neighboring variables in $G$~\citep{Zhou2016ArchimedesOne}. Specifically, consider the subgraph consisting of $v_i$ and its $r$-hop neighbors. It has been shown that increasing the diameter of neighborhood (the value of $r$) can effectively improve the approximation accuracy, and with even a small value of $r$ (e.g. 2-3), the approximation by $r$-hop inference can be sufficiently accurate in many real scenarios.
	
	 However, in the scenario of gradual inference, some factors (e.g. attribute value similarity) are usually shared by almost all the variables. As a result, the simple approach of considering $r$-hop neighborhood may result in a subgraph covering almost all the variables. Therefore, we propose to limit the size of inference subgraph in the following manner:
\begin{enumerate}
   \item Gradual learning infers the label of a pair based on its features. Approximate factor graph inference only needs to consider the factors corresponding to the features of $v_i$. The other factors in $G$ are instead excluded from the constructed subgraph;
   \item The influence distribution of a factor is estimated based on its evidence variables. Approximate factor graph inference only needs to consider the evidence variables sharing at least one feature with the target inference variable, $v_i$. The remaining variables, including the unlabeled inference variables other than $v_i$ and the evidence variables not sharing any common feature with $v_i$, are instead excluded from the constructed subgraph;
   \item In the case that applying the previous two guides still results in an exceedingly large subgraph, we propose to limit the total number of evidence variables for any given feature. As pointed out in the literature~\citet{Chen1994Empirical}, the accuracy of function regression generally increases with the number of sample observations. However, the validity of this proposition depends on the uniform distribution of the samples. The additional samples very similar to the existing ones can only produce marginal improvement on prediction accuracy. Therefore, we also limit the total number of evidence variables for any given feature. In practical implementation, we suggest to divide the feature value range of [0,1] into ten uniform intervals, [0,0.1], [0.1,0.2], $\ldots$, [0.9,1.0], and limit the number of observations for each interval (e.g. between 50 and 200).
\end{enumerate}		
	
	 It is worthy to point out that our proposed approach for subgraph construction is consistent with the principle of $r$-hop approximation in that it essentially opts to include those factors and variables in the close neighborhood of a target variable in the subgraph.

\section{Empirical Evaluation}  \label{sec:experiments}

    In this section, we empirically evaluate the performance of our proposed approach (denoted by GML) on real data. We compare GML with both unsupervised and supervised alternative techniques, which include
\begin{itemize}
  \item Unsupervised Clustering (denoted by UC). The approach of unsupervised clustering maps the record pairs to points in a multi-dimensional feature space and then clusters them into distinct classes based on the distance between them. The features usually include different similarity metrics specified at different attributes. In our implementation, we use the classical k-means technique to classify pairs into two classes.
	\item Unsupervised Rule-based (denoted by UR). The unsupervised rule-based approach reasons about pair equivalence based on the rules handcrafted by the human. Based on human experience and knowledge on the test data, the rules are specified in terms of record similarity. For fair comparison, in our implementation, UR first uses the result of unsupervised clustering (UC) to estimate the proportions of matching and unmatching instances in a workload, and then proportionally identify the easy matching and unmatching instances by record similarity.
  \item Learning based on Support Vector Machine (denoted by SVM). The SVM-based approach~\citep{christen2008automatic} also maps the record pairs to points in a multi-dimensional feature space. Unlike unsupervised clustering, it fits an optimal SVM classifier on labeled training data and then uses the trained model to label the pairs in the test data.
  \item Deep Learning (denoted by DNN). The deep learning approach~\citep{mudgal18deep} is the state-of-the-art supervised learning approach for ER. Representing each record pair by vector, it first trains a deep neural network (DNN) on labeled training data, and then uses the trained DNN to classify the pairs in the test data.
\end{itemize}			

  It is noteworthy that the existing semi-supervised learning and active learning techniques are usually applied in the scenario where only a limited number of labeled training data are available. Provided with enough training data, the performance of supervised techniques (e.g. DNN) can be expected to be no worse than their semi-supervised or active learning counterparts. Therefore, the aforementioned four techniques can provide a good coverage of the existing solutions for ER.

	The rest of this section is organized as follows: Subsection~\ref{sec:setup} describes the experimental setup. Subsection~\ref{sec:comparison} compares GML with the other alternatives. Subsection~\ref{sec:easylabeling} evaluates the sensitivity of GML w.r.t various parameter settings. Finally, subsection~\ref{sec:scalability} evaluates the scalability of GML.

\subsection{Experimental Setup} \label{sec:setup}

   Our evaluation is conducted on three real datasets, which are described as follows:
\begin{itemize}
\item DBLP-Scholar\footnote{available at https://dbs.uni-leipzig.de/file/DBLP-Scholar.zip} (denoted by DS): The DS dataset contains the publication entities from DBLP and the publication entities from Google Scholar. The experiments match the DBLP entries with the Scholar entries.
\item Abt-Buy\footnote{available at https://dbs.uni-leipzig.de/file/Abt-Buy.zip} (denoted by AB): The AB dataset contains the product entities from both Abt.com and Buy.com. The experiments match the Abt entries with the Buy entries.
\item Songs\footnote{available at http://pages.cs.wisc.edu/\~{}anhai/data/falcon\_data/songs} (denoted by SG): The SG dataset contains song entities, some of which refer to the same songs. The experiments match the song entries in the same table.
\end{itemize}

\begin{table*}[!h]
\small
\centering
\caption{Comparative Evaluation of GML}
\label{tb:compare}
\setlength{\tabcolsep}{2.83mm}{
\begin{tabular}{@{}c|ccc|ccc|ccc@{}}
\toprule
\multicolumn{1}{l|}{\multirow{2}{*}{}} & \multicolumn{3}{c|}{GML}         & \multicolumn{3}{c|}{UR}              & \multicolumn{3}{c}{UC}      \\ \midrule
\multicolumn{1}{l|}{}                  & recall & precision & F1          & recall   & precision  & F1           & recall  & precision & F1    \\ \midrule
DS                                     & 0.884  & 0.933     & {\bf 0.908} & 0.808    & 0.958      & {\bf 0.877}  & 0.793   & 0.939     & {\bf 0.860} \\
AB                                     & 0.632  & 0.546     & {\bf 0.586} & 0.773    & 0.300      & {\bf 0.432}  & 0.800   & 0.311     & {\bf 0.448} \\
SG                                     & 0.992  & 0.911     & {\bf 0.950} & 0.994    & 0.811      & {\bf 0.893}  & 0.995   & 0.808     & {\bf 0.892} \\ \bottomrule
\end{tabular}
}
\setlength{\tabcolsep}{2.68mm}{
\begin{tabular}{@{}c|ccc|ccc|ccc@{}}
\toprule
                        & \multicolumn{9}{c}{SVM}                                                             \\ \midrule
                        & \multicolumn{3}{c|}{10\%}   & \multicolumn{3}{c|}{20\%}   & \multicolumn{3}{c}{30\%}\\ \midrule
                        & recall & precision & F1              & recall & precision & F1              & recall & precision & F1         \\ \midrule
\multicolumn{1}{c|}{DS} & 0.890  & 0.918     & {\bf 0.903}     & 0.892  & 0.918     & {\bf 0.904}     & 0.896  & 0.921     & {\bf 0.908}\\
\multicolumn{1}{c|}{AB} & 0.418  & 0.771     & {\bf 0.527}     & 0.440  & 0.659     & {\bf 0.528}     & 0.423  & 0.700     & {\bf 0.528}\\
\multicolumn{1}{c|}{SG} & 0.995  & 0.855     & {\bf 0.920}     & 0.994  & 0.881     & {\bf 0.934}     & 0.994  & 0.892     & {\bf 0.940}\\ \bottomrule
\end{tabular}
}
\setlength{\tabcolsep}{2.65mm}{
\begin{tabular}{@{}c|ccc|ccc|ccc@{}}
\toprule
                        & \multicolumn{9}{c}{DNN}                                                                                      \\ \midrule
                        & \multicolumn{3}{c|}{10\%(5\%:5\%)} & \multicolumn{3}{c|}{20\%(15\%:5\%)} & \multicolumn{3}{c}{30\%(25\%:5\%)}\\ \midrule
                        & recall    & precision   & F1            & recall    & precision & F1            & recall    & precision    & F1\\ \midrule
\multicolumn{1}{c|}{DS} & 0.949     & 0.869       & {\bf 0.907}   & 0.945     & 0.956     & {\bf 0.950}   & 0.982     & 0.929        & {\bf 0.955}\\
\multicolumn{1}{c|}{AB} & 0.043     & 0.254       & {\bf 0.074}   & 0.441     & 0.601     & {\bf 0.509}   & 0.444     & 0.707        & {\bf 0.546}\\
\multicolumn{1}{c|}{SG} & 0.777     & 0.830       & {\bf 0.802}   & 0.952     & 0.900     & {\bf 0.925}   & 0.938     & 0.970        & {\bf 0.954}\\ \bottomrule
\end{tabular}
}
\end{table*}

 In the empirical study, GML uses pair similarity as the machine metric to identify easy instances. For fair comparison, given a percentage of easy instances (e.g. 30\%), GML first uses the result of unsupervised clustering (UC) to estimate the proportions of matching and unmatching instances in a workload, and then proportionally identify the easy matching and unmatching instances by record similarity. Pair similarity is computed by aggregating the attribute similarities via a weighted sum \citep{christen2012data}. Specifically, on the DS dataset, Jaccard similarity of the attributes \emph{title}, \emph{authors} and \emph{year}, and Jaro-Winkler distance of the attribute \emph{title}, \emph{authors} and \emph{venue} are used; on the AB dataset, Jaccard similarity of the attributes \emph{product name} and \emph{product description} are used; on the SG dataset, Jaccard similarity of the attributes \emph{song title}, Jaro-Winkler distance of the attributes \emph{song title} and \emph{release information}, and number similarity of the attributes \emph{duration} are used. The weight of each attribute is determined by the number of its distinct values. As in the previous study~\citep{mudgal18deep}, we use the blocking technique to filter the instance pairs having a small chance to be equivalent. After blocking, the DS workload has 10482 pairs, and 4771 among them are equivalent; the AB workload has 8924 pairs, and 774 among them are equivalent; the SG workload has 8312 pairs, and 1412 among them are equivalent. Our implementation codes of GML and the used test datasets have been made open-source available at our website\footnote{http://www.wowbigdata.com.cn/GML/GML.html}.

\subsection{Comparative Study} \label{sec:comparison}

	
   This section compares GML with its alternatives. In the comparative study, we set the ratio of easy instances at 30\% in all the experiments. For scalable gradual inference, we set $m=2000$ and $k=10$. Our evaluation results in Subsection.~\ref{sec:easylabeling} will show that GML can perform robustly w.r.t various parameter settings.

	The detailed evaluation results are presented in Table~\ref{tb:compare}, in which the results on F-1 have been highlighted. For the supervised approaches of SVM and DNN, we report their performance provided with different sizes of training data, which is measured by the fraction of training data among the whole dataset. In Table~\ref{tb:compare}, the percentage of training data is listed at the second low in the table. For instance, for SVM, ``30\%'' means that 30\% of a dataset are used for training and the remaining 70\% are test data. For DNN, the training data consists of the data used for model training and the data used for validation. Therefore, we report the fractions of both parts in the table. For instance, ``30\%(25\%:5\%)'' means that 25\% of a dataset are used for training, 5\% are used for verification, and the remaining 70\% are test data. In the empirical evaluation, training data are randomly selected from the workload. Since the performance of SVM and DNN depends on the randomly-selected training data, the reported results are the averages over ten runs.

	The results show that GML performs considerably better than the unsupervised alternatives, UR and UC. In most cases, their performance differences on F-1 are larger than 5\%. The performance of GML in terms of F-1 is also highly competitive compared to both supervised approaches of SVM and DNN. It can be observed that GML can beat both supervised approaches of SVM and DNN in most cases if the percentage of provided training data is less than 20\%. When the size of training data increases, the performance of SVM and DNN generally improves as expected. Even with the training data size at 30\%, GML beats SVM on AB and SG while achieving roughly the same performance as SVM on DS. GML also beats DNN by more than 4\% on AB while achieving roughly the same performance as DNN on SG. \emph{It is worthy to point out that unlike the supervised SVM and DNN models, GML does not use any labeled training data. These experimental results evidently demonstrate the efficacy of GML.}

\subsection{Sensitivity Evaluation} \label{sec:easylabeling}

   \begin{table}[!h]
\small
\centering
\caption{Sensitivity Evaluation w.r.t Easy Instance Labeling}
\label{tb:easyproportion}
\setlength{\tabcolsep}{11.60mm}{
\begin{tabular}{@{}c|c|c|l@{}}
\toprule
\multicolumn{1}{c|}{F-1} & \multicolumn{1}{c|}{$30\%$} & \multicolumn{1}{c|}{$40\%$}  & \multicolumn{1}{c}{$50\%$}  \\ \midrule
\multicolumn{1}{c|}{DS}  & 0.908                      & 0.908                       & 0.909~~~~                     \\
\multicolumn{1}{c|}{AB}  & 0.586                      & 0.560                       & 0.540~~~~                     \\
\multicolumn{1}{c|}{SG}  & 0.950                      & 0.950                       & 0.950~~~~                     \\ \bottomrule
\end{tabular}}
\end{table}

\begin{table}[!h]
\small
\centering
\caption{Sensitivity Evaluation w.r.t the Parameter $m$}
\label{tb:m}
\setlength{\tabcolsep}{9.00mm}{
\begin{tabular}{@{}c|c|c|l@{}}
\toprule
\multicolumn{1}{c|}{F-1} & \multicolumn{1}{c|}{$m=500$} & \multicolumn{1}{c|}{$m=1000$} & \multicolumn{1}{c}{$m=2000$}\\ \midrule
\multicolumn{1}{c|}{DS}  & 0.906                        & 0.906                         & ~~~0.908                       \\
\multicolumn{1}{c|}{AB}  & 0.572                        & 0.576                         & ~~~0.586                       \\
\multicolumn{1}{c|}{SG}  & 0.950                        & 0.950                         & ~~~0.950                       \\ \bottomrule
\end{tabular}
}
\end{table}

\begin{table}[!h]
\small
\centering
\caption{Sensitivity Evaluation w.r.t the Parameter $k$}
\label{tb:k}
\setlength{\tabcolsep}{10.80mm}{
\begin{tabular}{@{}c|c|c|l@{}}
\toprule
\multicolumn{1}{c|}{F-1} & \multicolumn{1}{c|}{$k=1$} & \multicolumn{1}{c|}{$k=5$}  & \multicolumn{1}{c}{$k=10$}  \\ \midrule
\multicolumn{1}{c|}{DS}  & 0.906                      & 0.906                       & ~0.908                      \\
\multicolumn{1}{c|}{AB}  & 0.577                      & 0.581                       & ~0.586                      \\
\multicolumn{1}{c|}{SG}  & 0.950                      & 0.950                       & ~0.950                      \\ \bottomrule
\end{tabular}}
\end{table}

\begin{table}[!h]
\small
\centering
\caption{Sensitivity Evaluation w.r.t the Parameter $\delta$}
\label{tb:u}
\setlength{\tabcolsep}{10.00mm}{
\begin{tabular}{@{}c|c|c|l@{}}
\toprule
\multicolumn{1}{c|}{F-1} & \multicolumn{1}{c|}{$\delta=50$} & \multicolumn{1}{c|}{$\delta=100$} & \multicolumn{1}{c}{$\delta=200$}\\ \midrule
\multicolumn{1}{c|}{DS}  & 0.906                      & 0.906                       & ~0.908                      \\
\multicolumn{1}{c|}{AB}  & 0.578                      & 0.582                       & ~0.586                      \\
\multicolumn{1}{c|}{SG}  & 0.950                      & 0.950                       & ~0.950                      \\
\bottomrule
\end{tabular}}
\end{table}

   In this section, we evaluate the sensitivity of GML w.r.t different parameter settings. We first vary the ratio of the initial easy instances in a workload and track the performance of GML with different ratios. For scalable gradual inference, we vary the number of the pair candidates selected for inference probability approximation (the parameter $m$ in Algorithm~\ref{alg:gradualinference}), the number of the pair candidates selected for factor graph inference (the parameter $k$ in Algorithm~\ref{alg:gradualinference}), and the limit specification on the number of evidential observations for each interval of feature value (i.e. [x,x+0.1]) (the parameter $\delta$) in the constructed inference subgraph. The value of $m$ is set between 500 and 2000, the value of $k$ is set between 1 and 10 and the value of $\delta$ is set between 50 to 200. While evaluating the sensitivity of GML w.r.t a specific parameter, we fixed all the other parameters at the same values.
	
	The detailed evaluation results w.r.t various parameter settings are reported in Table~\ref{tb:easyproportion}, ~\ref{tb:m}, ~\ref{tb:k} and ~\ref{tb:u}. We can see that given a reasonable range on the ratio of easy instances (between 30\% and 50\%), the performance of GML is stable. On DS and SG, the performance of GML only fluctuates very marginally. On AB, the performance of GML deteriorates slightly as the ratio is set higher. Our closer scrutiny reveals that on AB, the accuracy of easy instance labeling decreases as the ratio increases from 30\% to 50\%. As a result, the performance of GML deteriorates accordingly, However, with the ratio set at 50\%, the performance of GML (0.540 measured by F-1) is still competitive compared to SVM and DNN. Similarly, as shown Table~\ref{tb:m}, ~\ref{tb:k} and ~\ref{tb:u}, the performance of GML is highly robust w.r.t the parameters of $m$, $k$ and $\delta$. Our experimental results bode well for GML's applicability in real applications.
	


	
	It is worthy to point out that even though setting $k$ to a small number can only marginally affect the performance of GML, it does not mean that the factor graph inference is unwanted, can thus be replaced by the more efficient approximate probability estimation. On the contrary, in the experiments, we have observed that there actually exist many pair instances whose factor graph inference results are sufficiently different from their approximated probabilities such that their labels are flipped by factor graph inference, especially in the final stages of gradual inference.

\subsection{Scalability Evaluation} \label{sec:scalability}

\begin{figure}[h!]
\centering
\includegraphics[width=0.8\linewidth]{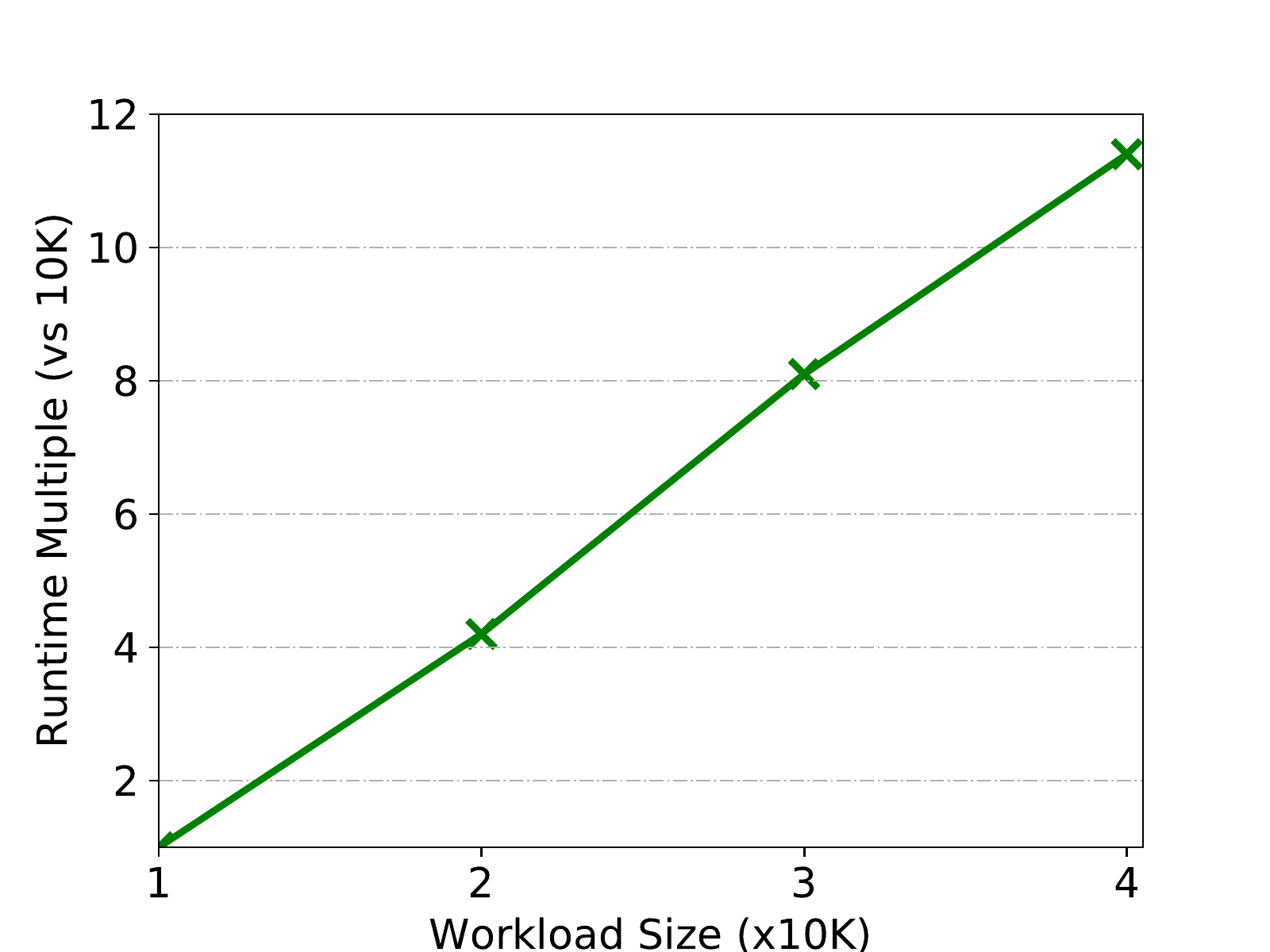}
\caption{Scalability Evaluation.}
\label{fig:scalability}
\end{figure}

   In this section, we evaluate the scalability of the proposed scalable approach for GML. Based on the entities in DBLP and Scholar, we generate different-sized DS workloads, from 10000 to 40000. We fix the proportion of identified easy instances at 50\%, the value of $m$ at 2000, the value of $k$ at 1 and the value of $\delta$ at 50. The detailed evaluation results on scalability are presented in Figure~\ref{fig:scalability}, in which the x-axis denotes workload size and the y-axis denotes the cost multiple with the runtime spent on the workload of $10k$ as the baseline. In the experiments, we have observed that even though the total number of features consistently increases with workload size, the number of features any instance has is quite stable (in the order of tens). Because the number of evidential observations for each interval of feature values is limited by $\delta$, the average cost of the scalable GML spent on each unlabeled pair only increases marginally as the workload increases. As a result, the total consumed time increases nearly linearly with workload size. Our experimental results clearly demonstrate that the proposed scalable approach scales well with workload size.

\section{Conclusion} \label{sec:conclusion}

   In this paper, we have proposed a novel learning paradigm, called gradual machine learning, which begins with some easy instances in a given task, and then gradually labels more challenging instances in the task based on iterative factor graph inference without requiring any human intervention. We have also developed an effective solution based on it for the task of entity resolution. Our extensive empirical study on real data has shown that the performance of the proposed solution is considerably better than the unsupervised alternatives, and highly competitive compared to the state-of-the-art supervised techniques. Using ER as a test case, we have demonstrated that gradual machine learning is a promising paradigm potentially applicable to other challenging classification tasks requiring extensive labeling effort.
	
	 Our research on gradual machine learning is an ongoing effort. For future work, even though gradual machine learning is proposed as an unsupervised learning paradigm in this paper, human work can be potentially integrated into its process for improved performance. An interesting open challenge is then how to effectively improve the performance of gradual machine learning with the minimal effort of human intervention, which include but are not limited to manually labeling some instances. On the other hand, it is very interesting to develop the solutions based on the propose paradigm for other challenging classification tasks besides entity resolution and sentiment analysis.

\vskip 0.2in
\bibliographystyle{plain}

\end{document}